# Nonlinear Processes Responsible for mid-Infrared and Blue Light Generation in Alkali Vapours

Alexander Akulshin[1], Dmitry Budker[2], Brian Patton[2], and Russell McLean[1]

[1]Centre for Atom Optics and Ultrafast Spectroscopy,
Swinburne University of Technology, PO Box 218, Melbourne 3122, Australia
[2]Department of Physics, University of California, Berkeley, CA 94720-7300, USA
E-mail: aakoulchine@swin.edu.au

**Abstract:** The nonlinear processes responsible for frequency up- and down-conversion of resonant low-intensity laser radiation in Rb vapour have been evaluated from the spatial and temporal properties of blue and mid-IR light resulting from wave mixing.

## 1. Introduction

Nonlinear parametric and nonparametric processes in atomic media can produce new optical fields with substantially different optical frequencies. Frequency conversion of low-intensity cw laser radiation of diode lasers into highly directional blue and mid-IR light in Rb and Cs vapours [1-4] is an active area of research because of potential applications in quantum information processing, underwater communication and remote-detection magnetometry [5].

In this approach, the new field generation occurs without an optical cavity. Alkali vapours provide not only high resonant nonlinearity, but also set strong spatial anisotropy for new-field generation. An atomic medium driven by bi-chromatic laser radiation tuned close to ladder-type transitions, $5S_{1/2} \rightarrow 5P_{3/2}$ and $5P_{3/2} \rightarrow 5D_{5/2}$ in the case of Rb atoms, can produce superfluorescent emission at 5.23 μm, which is resonant to the $5D_{5/2} \rightarrow 6P_{3/2}$ transition. This mid-IR radiation can even be randomly directed in a dense atomic sample [6]; however, in the case of modest atomic density (N ≤ 1×10$^{13}$ cm$^{-3}$) and a pencil-shaped atom-light interaction region, superfluorescence consists of collimated forward and backward radiation. Mixing of the forward superfluorescence mode with the applied laser fields produces radiation at 420 nm in the co-propagating direction only, satisfying the phase-matching relation: $\boldsymbol{k}_1+\boldsymbol{k}_2=\boldsymbol{k}_{IR}+\boldsymbol{k}_{BL}$, where $\boldsymbol{k}_{1,2}$ are the wave vectors of the laser fields at 780 and 776 nm, while $\boldsymbol{k}_{IR}$ and $\boldsymbol{k}_{BL}$ are the wave vectors at 5.23 μm and 420 nm radiation, respectively.

## 2. Experimental results

In our studies the spectral width of the collimated blue light (CBL) has been found to be less than 2 MHz for a wide range of experimental conditions [7]. The linewidth appears to be mainly determined by the spectral properties of the laser radiation rather than the parameters of the atomic medium. The linewidth and optical frequency of CBL remain unchanged to within the 0.5 MHz experimental resolution, despite at least two-fold variations in the atomic density and in the intensity of each laser. The CBL frequency is found to be centered on the $5S_{1/2}(F=3) \rightarrow 6P_{3/2}(F'=4)$ transition, and can be tuned over a range of more than 250 MHz by tuning the laser frequencies.

The effect of velocity selective hyperfine optical pumping on amplified spontaneous emission and parametric wave mixing has been explored. The number of resonant atoms contributing to the nonlinear processes can be increased or decreased by velocity-selective incoherent optical pumping produced by an additional laser tuned to any open transition from the ground-state $5S_{1/2}$ level. For a wide range of atomic densities ($3 \times 10^{10}$ cm$^{-3}$ < N < $3 \times 10^{11}$ cm$^{-3}$) at least tenfold enhancement of the coherent and directional blue radiation due to optical pumping has been observed. The parametric wave mixing in optically pumped Rb vapours has been obtained at a cell temperature as low as 33.6 $^0$C, corresponding to an atomic density of $2.8 \times 10^{10}$ cm$^{-3}$.

It has been shown that optical pumping also modifies the refractive index of the medium, perturbing the phase matching condition which must be satisfied and, consequently, affecting the direction of the CBL. Thus, parameters of co-propagating blue light, such as intensity, direction and divergence, can be controlled by optical pumping and such control is an important step towards probing small numbers of atoms using this approach.

We have undertaken an experimental study of the spectral and spatial properties of both the mid-IR and blue radiation generated in specially fabricated Rb vapour cells. In order to increase the atom-light interaction time and the probability of velocity-changing collisions we have extended our investigations to atomic ensembles contained in cells with anti-relaxation internal surface coating and a buffer gas. A cell with paraffin anti-relaxation coating allows effects related to hyperfine and Zeeman optical pumping produced by an additional laser to be emphasized. In our previous studies it was not possible to detect the mid-IR radiation directly, and its characteristics had to be inferred, but in a cell with sapphire windows transparent for IR radiation both the optical fields can be directly detected, as shown in Fig.1. Different spectral dependences of the blue and mid-IR radiation reflect their different origin.

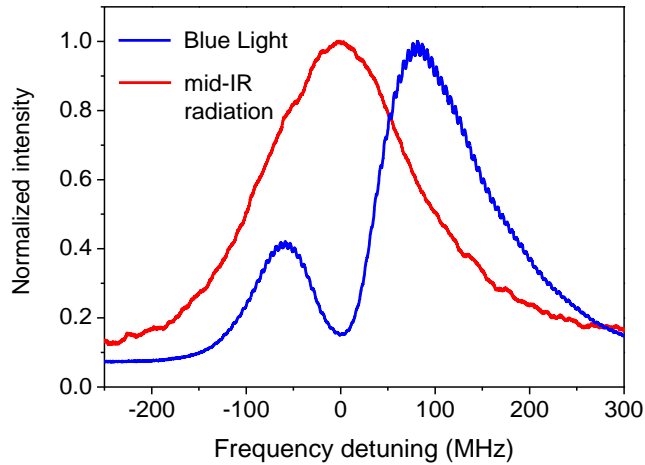

Fig. 1. Normalized intensity of co-propagating blue and mid-IR radiation as a function of the frequency offset of the 776 nm laser from the two-photon transition at fixed frequency of the 780 nm laser locked to the $5S_{1/2}(F=3) \rightarrow 5P_{3/2}(F'=4)$ transition.

We have found that the divergence of the co-propagating blue and mid-IR light is similar (approximately 6 mrad) and determined by the collimation of the applied laser light. The backward mid-IR radiation is weaker and more divergent (~14 mrad). Also the forward and backward mid-IR superfluorescence have different dependences on the laser detunings.

## 3. Conclusion

There is substantial interest in enhancing efficiency of laser remote sensing. Backward directional and efficient emission may constitute a novel approach to the problem. Further investigation of this and other nonlinear schemes based on resonant wave mixing and understanding the role of various parametric and nonparametric processes in dilute atomic samples should also determine the usefulness of this idea for laser-guide-star techniques [8]. This scheme could also be used for generation of coherent and correlated fields at wavelengths that are difficult to access with other methods.

## 4. References

[1] A. S. Zibrov, M. D. Lukin, L. Hollberg, and M. O. Scully, "Efficient frequency up-conversion in resonant coherent media", Phys. Rev. A **65**(5), 051801 (2002).

[2] A. M. Akulshin, R. J. McLean, A. I. Sidorov, and P. Hannaford, "Coherent and collimated blue light generated by four-wave mixing in Rb vapour", Optics Express, **17**(25) 22861, (2009).

[3] J. T. Schultz, S. Abend, D. Döring, J. E. Debs, P. A. Altin, J. D. White, N. P. Robins, and J. D. Close, "Coherent 455 nm beam production in a cesium vapor", Opt. Lett. **34**, 2321 (2009).

[4] G. Walker, A.S. Arnold, S. Franke-Arnold, "Trans-Spectral Orbital Angular Momentum Transfer via Four-Wave Mixing in Rb Vapor", Phys. Rev. Lett. **108**, 243601 (2012).

[5] J. M. Higbie, S. M. Rochester, B. Patton, R. Holzlöhner, D. B. Calia, and D. Budker, "Magnetometry with mesospheric sodium", Proc. Nat. Acad. Science **108**, 3522 (2011).

[6] A. I. Lvovsky, S. R. Hartmann and F. Moshary, "Omnidirectional superfluorescence", Phys. Rev. Lett. **82**, 4420 (1999).

[7] A. Akulshin, Ch. Perrella, G-W. Truong, R. McLean, and A. Luiten, "Frequency evaluation of collimated blue light generated by wave mixing in Rb vapour", J. Phys. B **45**, 245503 (2012).

[8] W. Happer, G. J, MacDonald, C. E. Max, F. J. J. Dyson, "Atmospheric-turbulence compensation by resonant optical backscattering from the sodium layer in the upper atmosphere", Opt. Soc. Am. A**11**, 263 (1994).